\documentclass{sig-alternate-10pt}
\pdfoutput=1
\usepackage[T1]{fontenc}
\usepackage{times}  
\usepackage{epsfig}
\usepackage{afterpage}
\usepackage{tabularx}
\usepackage{graphicx}
\usepackage{balance}
\usepackage{color}
\usepackage{xcolor}
\usepackage{xspace}
\usepackage{thumbpdf}
\usepackage{listings}
\usepackage{verbatim}
\usepackage{color}
\usepackage[hidelinks]{hyperref}
\definecolor{darkred}{rgb}{0.7,0,0}
\definecolor{darkgreen}{rgb}{0,0.5,0}
\hypersetup{colorlinks=true,
        linkcolor=darkred,
        citecolor=darkgreen}
\usepackage{booktabs}
\usepackage{colortbl}
\usepackage[inline]{aplcomments}
\usepackage{inconsolata}
\usepackage{paralist}
\usepackage{xspace}
\usepackage{listings}
\usepackage{breakurl}
\usepackage{inconsolata}
\usepackage{longtable}
\usepackage{placeins}
\usepackage{caption, subcaption}
\usepackage{pifont}% http://ctan.org/pkg/pifont
\usepackage{tablefootnote}
\usepackage{siunitx}
\newcommenter{ak}{1.0,1.0,0.3}
\newcommenter{ac}{0.4,1.0,1.0}
\newcommand{\pktlanguage}{Domino\xspace}
\newcommand{\absmachine}{Banzai\xspace}

\lstdefinestyle{customc}{
 belowcaptionskip=1\baselineskip,
 breaklines=true,
 xleftmargin=20pt,
 language=C,
 frame=L,
 escapeinside={@}{@},
 showstringspaces=false,
 basicstyle=\small\ttfamily,
 keywordstyle=\bfseries\color{green!40!black},
 commentstyle=\itshape\color{purple!40!black},
 stringstyle=\color{orange},
 directivestyle=\color{brown},
 numbers=left, numberstyle=\tiny\color{gray}
}

\lstdefinestyle{customctable}{
 aboveskip=-\medskipamount,
 belowskip=-\medskipamount,
 language=C,
 escapeinside={@}{@},
 showstringspaces=false,
 basicstyle=\scriptsize\ttfamily,
 keywordstyle=\bfseries\color{green!40!black},
 commentstyle=\itshape\color{purple!40!black},
 stringstyle=\color{orange},
 directivestyle=\color{brown},
}

\def\compactify{\itemsep=0pt \topsep=0pt \partopsep=0pt \parsep=0pt}
\let\latexusecounter=\usecounter

\newenvironment{CompactEnumerate}
  {\def\usecounter{\compactify\latexusecounter}
   \begin{enumerate}}
  {\end{enumerate}\let\usecounter=\latexusecounter}

\newcommand{\MA}[1]{}
\newcommand{\hb}[1]{}

\sloppypar
\begin{document}

%don't want date printed
\date{}

%make title bold and 14 pt font (Latex default is non-bold, 16 pt)
%\title{\Large \bf Wonderful : A Terrific Application and Fascinating Paper}
%\title{Packet Transactions: Programming Data-Plane Algorithms at Line Rate}
\title{Packet Transactions: High-level Programming for Line-Rate Switches}
\author{
  \fontsize{10.7}{9.9}\rm Anirudh Sivaraman\textsuperscript{*}, Mihai Budiu\textsuperscript{\dag}, Alvin Cheung\textsuperscript{\ddag}, Changhoon Kim\textsuperscript{\dag}, Steve Licking\textsuperscript{\dag}, \\
  \fontsize{10.7}{9.9}\rm George Varghese\textsuperscript{++}, Hari Balakrishnan\textsuperscript{*}, Mohammad Alizadeh\textsuperscript{*}, Nick McKeown\textsuperscript{+}\\
\fontsize{10.7}{9.9}\selectfont \textsuperscript{*}MIT CSAIL, \textsuperscript{\dag}Barefoot Networks, \textsuperscript{\ddag}University of Washington, \textsuperscript{++}Microsoft Research, \textsuperscript{+}Stanford University
}
% copy the following lines to add more authors
% \and
% {\rm Name}\\
%Name Institution
%} % end author

\maketitle

% Use the following at camera-ready time to suppress page numbers.
% Comment it out when you first submit the paper for review.
%\thispagestyle{empty}
\begin{abstract}

Many algorithms for congestion control, scheduling, network measurement,
active queue management, security, and load balancing require custom processing
of packets as they traverse the data plane of a network switch. To run at
line rate, these data-plane algorithms must be in hardware. With today's switch
hardware, algorithms cannot be changed, nor new algorithms installed, after a
switch has been built.

This paper shows how to program data-plane algorithms in a high-level language
and compile those programs into low-level microcode that can run on emerging
programmable line-rate switching chipsets. The key challenge is that these
algorithms create and modify algorithmic state. The key idea to achieve
line-rate programmability for stateful algorithms is the notion of a {\em
packet transaction}: a sequential code block that is atomic and isolated from
other such code blocks. We have developed this idea in \pktlanguage, a C-like
imperative language to express data-plane algorithms. We show with many
examples that \pktlanguage provides a convenient and natural way to express
sophisticated data-plane algorithms, and show that these algorithms can be run
at line rate with modest estimated die-area overhead.

\end{abstract}

\section{Introduction}
\label{s:intro}

Network switches and routers in modern datacenters, enterprises, and
service-provider networks perform many tasks in addition to standard
packet forwarding. The set of requirements for routers has only
increased with time as network operators seek greater control over
performance and security.  Performance and security may be improved
using both data-plane and control-plane mechanisms. This paper focuses
on data-plane algorithms. These algorithms process and transform
packets, creating and maintaining state in the switch. Examples
include active queue management~\cite{red,blue,avq,codel,pie},
scheduling~\cite{pifo_hotnets}, congestion control with switch
feedback~\cite{xcp, rcp, pdq, dctcp}, network
measurement~\cite{opensketch, bitmap_george, elephant_george},
security~\cite{dns_change}, and traffic load balancing~\cite{conga}.

An important requirement for data-plane algorithms is the ability to
process packets at the switch's line rate (typically 10--100 Gbit/s on
10--100 ports).  As a result, these
algorithms are typically implemented using dedicated
hardware. Hardware designs are rigid, however, and not reconfigurable
in the field. Thus, to implement and deploy a new algorithm today, or
to even modify a deployed one, the user must invest in new
hardware---a time-consuming and expensive proposition.

This rigidity affects many stakeholders adversely:
vendors~\cite{cisco_nexus, dell_force10, arista_7050} building network
switches with merchant-silicon chips~\cite{trident, tomahawk,
  mellanox}, network operators deploying
switches~\cite{google,facebook,vl2}, and researchers developing new
switch algorithms~\cite{xcp, codel, d3, detail, pdq}.  

To run data-plane algorithms after a switch has been built,
researchers and companies have attempted to build programmable routers
for many years, starting from efforts on active
networks~\cite{active-nets} to network processors~\cite{npu_survey} to
software routers~\cite{click, dpdk, routebricks}. All these efforts
sacrificed performance for programmability, typically running an order
of magnitude (or worse) slower than hardware line
rates. Unfortunately, this reduction in performance has meant that
these systems are rarely deployed in production networks, if at all.

Programmable switching chips~\cite{flexpipe, xpliant, rmt, corsa,
  uadp, algo_logic} competitive in performance with state-of-the-art
fixed-function chipsets~\cite{trident, tomahawk, mellanox} are now
becoming available. These chips implement a few low-level hardware
primitives that can be configured by software into a processing
pipeline, and are
field-reconfigurable~\cite{xpliant_sdk,xpliant_sdk2,intel_sdk}. Building
a switch with such a chip is attractive because it does not compromise
on data rates~\cite{rmt}.

In terms of programmability, these chips today allow the network
operator to specify packet parsing and forwarding without restricting
the set of protocol formats or the set of actions that can be executed
when matching packet headers in a match-action table. Languages such
as P4 are emerging as a way to express such {\em match-action
  processing} in a hardware-independent way~\cite{p4,p4spec,dc_p4}.

There is a gap between this form of programmability and the needs of data-plane
algorithms. By contrast to packet header parsing and forwarding, which don't
modify state in the data plane, many data-plane algorithms create and modify
algorithmic state in the switch as part of packet processing. For such
algorithms, it is important for programmability to directly capture the
algorithm's intent without requiring it to be ``shoehorned'' into hardware
constructs such as a sequence of match-action tables. Indeed, this is how such
data-plane algorithms are expressed in pseudocode~\cite{red, csfq, codel_code,
avq, blue}, and implemented in software routers~\cite{click, dpdk,
routebricks}, network processors~\cite{packetc, nova}, and network
endpoints~\cite{qdisc}.

By studying the requirements of data-plane algorithms and the
constraints of line-rate hardware, we introduce a new abstraction to
program and implement data-plane algorithms: a {\em packet
  transaction} (\S\ref{s:transactions}). A packet transaction is a
sequential code block that is atomic and isolated from other such code
blocks (i.e., any visible state is equivalent to a serial execution of
packet transactions across packets). Packet transactions allow the programmer to
focus on the operations needed for each packet without worrying about
other concurrent packets.

We have designed and implemented {\em \pktlanguage{}}, a new
domain-specific language (DSL) for data-plane algorithms, with packet
transactions at its core.  \pktlanguage is an imperative language with
C-like syntax, perhaps the first to offer such a high level
programming abstraction for line-rate switches.

This paper makes three further contributions. First, {\em
  \absmachine}, a machine model for line-rate programmable switches
(\S\ref{s:absmachine}). \absmachine generalizes and abstracts
essential features of line-rate programmable switches~\cite{rmt,
  xpliant, flexpipe}. \absmachine also models practical constraints
limiting stateful operations at line rate.  Informed by these
constraints, we introduce the concept of {\em atoms} to represent a
programmable switch's instruction set.

Second, {\em a compiler from \pktlanguage packet transactions to a
  \absmachine target}~(\S\ref{s:compiler}). The \pktlanguage compiler
introduces \textit{all-or-nothing compilation}, where all packet
transactions accepted by the compiler will run at line rate, or be
rejected outright. There is no ``slippery slope'' of running network
algorithms at lower speeds as with traditional network processors or
software routers: when compiled, a \pktlanguage program runs at the
line rate, or not at all. Performance is not just predictable, but
is guaranteed.

Third, {\em an evaluation of \pktlanguage} (\S\ref{s:eval}). We evaluate
\pktlanguage's expressiveness by programming a variety of data-plane
algorithms (Table~\ref{tab:algos}) in \pktlanguage and compare with
P4. We find that \pktlanguage provides a more concise and easier
programming model for stateful data-plane algorithms.  Next, because
no existing programmable switch supports the set of atoms required for
our data-plane algorithms, we design a set of compiler targets
(\S\ref{ss:targets}) based on \absmachine and show that these are
feasible in a 32 nm standard-cell library with $< 15\%$ estimated chip area
overhead.  Finally, we compile data-plane algorithms written in
\pktlanguage to these targets to show how the choice of atoms in a
target determines which algorithms it can support.

\section{A Machine Model for Line-rate Switches}
\label{s:absmachine}

\begin{figure*}[!t]
  \includegraphics[width=\textwidth]{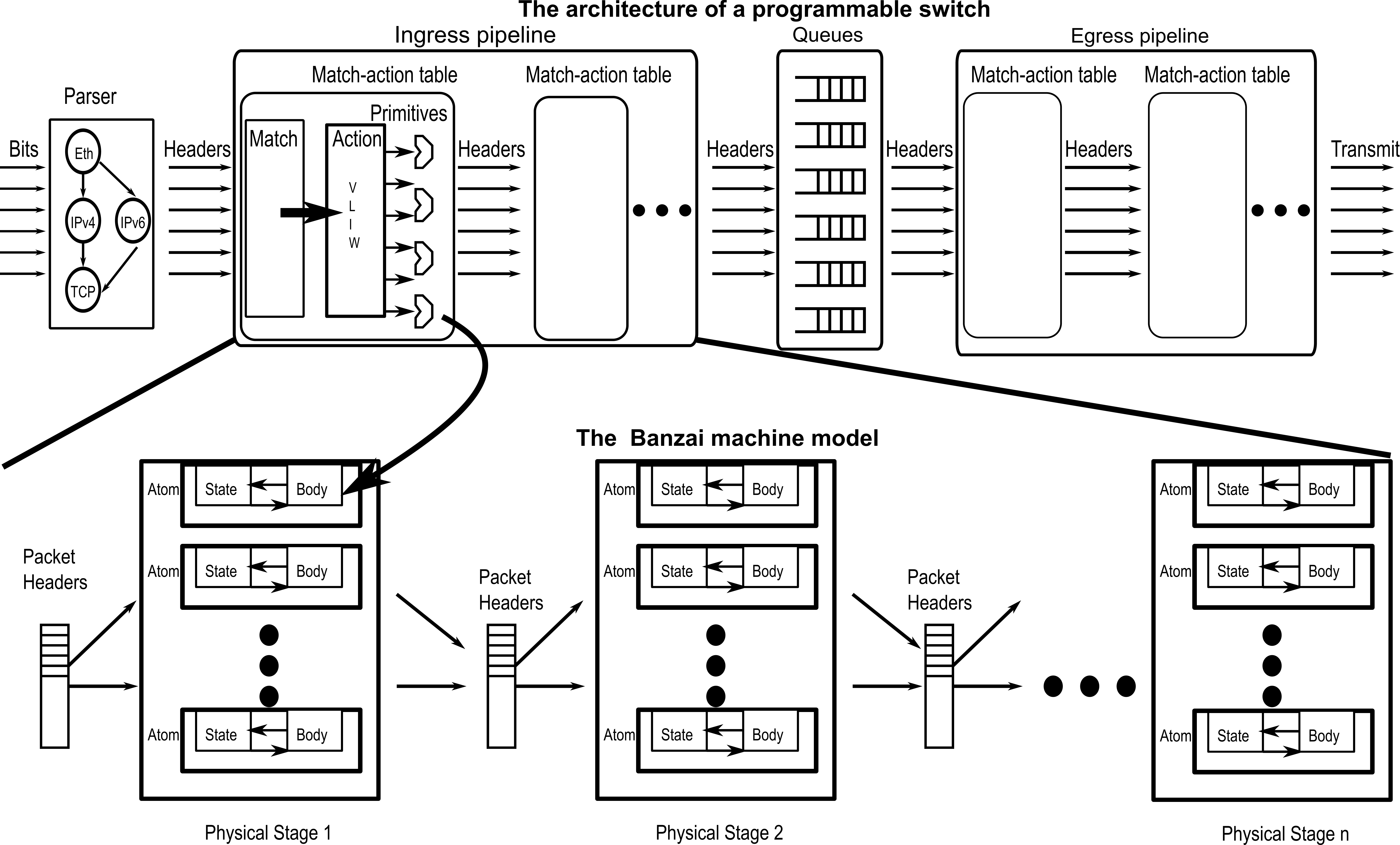}
  \caption{The \absmachine machine model and its relationship to
  programmable switch architectures.}
  \label{fig:switch}
\end{figure*}

\absmachine is a machine model for programmable line-rate switches
that serves as the compiler target for \pktlanguage programs.
\absmachine's design is inspired by recent programmable switch
architectures such as RMT~\cite{rmt}, Intel's
FlexPipe~\cite{flexpipe}, and Cavium's XPliant Packet
Architecture~\cite{xpliant}. \absmachine abstracts these architectures
and extends them with stateful processing units to implement
data-plane algorithms. These processing units, called {\em atoms},
precisely model the set of operations that a hardware target can
execute at line rate; they function as the target instruction set for
the \pktlanguage compiler.

\subsection{Background: Programmable switches}
Packets arriving at a programmable switch~(Figure~\ref{fig:switch}) are parsed
by a programmable parser that turns packets into header fields. These header
fields are first processed by an ingress pipeline consisting of match-action
tables arranged in stages. Processing a packet at a stage may modify its header
fields as well as some persistent state at that stage. Each stage has access
only to its own local state. To share state between stages, it must be carried
forward in packet headers. Following the ingress pipeline, the packet is
queued. Once the packet is dequeued by the switch scheduler, it is processed by
a similar egress pipeline before being transmitted.

To reduce chip area, the ingress and egress pipelines are shared across switch
ports.  Each pipeline handles aggregate traffic belonging to all ports on the
switch, at all packet sizes.  For instance, a 64-port switch with a line rate
of 10 Gbits/s per port and a minimum packet size of 64 bytes needs to process
around a billion packets per second~\cite{rmt}.  Equivalently, with a clock
frequency of 1 GHz, each pipeline stage needs to process one packet every clock
cycle (1 ns).  The need to handle one packet per clock cycle is typical because
switches are designed for the highest port count and line rate for a given chip
area. We assume one packet per clock cycle throughout the paper.\footnote{For
concreteness, we assume a 1 GHz clock frequency.}

Having to process a packet every clock cycle in each stage greatly
constrains the operations that can be performed on each packet. In
particular, any packet operation that modifies state visible to the
next packet {\em must} finish execution in a single clock cycle (see
\S\ref{ss:atoms} for details). Because of this restriction,
programmable switching chips provide a small set of processing units
or primitives for manipulating packets and state in a stage, unlike in
software routers. These processing units determine what algorithms can
run on the switch at line rate.

The challenge here is to determine primitives that allow a broad range of
data-plane algorithms to be implemented, and build a compiler to map a
user-friendly description of an algorithm to the primitives provided by a
switch.

\subsection{The \absmachine machine model}

\absmachine (the bottom half of Figure~\ref{fig:switch}) models the data-plane
components of an ingress or egress switch pipeline, consisting of a number of
stages executing synchronously on every clock cycle. Each stage processes one
packet every clock cycle (1 ns) and hands it off to the next, until it exits
the pipeline. \absmachine models the computation within a match-action table in
a stage (i.e., the action half of the match-action table), but not the match
semantics (e.g., direct, or ternary) (we discuss how to embed these
computations in a standard match-action pipeline in \S\ref{ss:guards}).
\absmachine does not model packet parsing and assumes that packets arriving to
it are already parsed.

\subsection{Atoms: \absmachine's processing units}
\label{ss:atoms}

Each pipeline stage in \absmachine contains a {\em vector of
  atoms}. All atoms in the vector execute in parallel on every clock
cycle.  Informally, an atom is an atomic unit of packet processing
supported natively by a \absmachine machine.
The atoms provided by 
a \absmachine machine form its instruction set.
Atoms may modify persistent state stored on the
switch. In contrast to instruction sets for CPUs, GPUs, DSPs, and
NPUs, the atoms for a \absmachine machine need to be substantially
richer to run real-world data-plane algorithms at line rate. We
explain why with an example.

Suppose we need to atomically increment a state variable stored on the switch
to count packets. One approach would be to have hardware support for three
simple single-cycle operations: \textit{read} some memory in the first clock
cycle, \textit{add} one in the next, and \textit{write} it to memory in the
third. This approach, however, does not provide atomic isolation. To see why,
suppose packet $A$ increments the counter from 0 to 1 by executing the read,
add, and write operations at clock cycles 1, 2, and 3 respectively.  If packet
$B$ issues the read at time 2, it will increment the counter again from 0 to 1,
when it should be 2. Locks over the shared counter are a potential
solution. However, locking causes packet $B$ to wait during packet $A$'s
increment, and the switch no longer sustains line rate of one packet every
clock cycle.\footnote{Wait-free objects~\cite{herlihy_wait} are an alternative
  to locking, but are typically too complex for hardware.} CPUs employ microarchitectural
  techniques such as operand forwarding to address this problem, but these techniques
  suffer from occasional pipeline stalls, which militates against line-rate
  performance.

The only way to provide an atomic increment is to explicitly support it in
hardware with an {\em atom} to read memory, increment it, and write it back in
a single stage within one clock cycle. The same observation applies to any
other line-rate atomic operation.

This observation motivates why we represent an atom as a body of sequential
code. An atom completes execution of the entire body of code and modifies a
packet before processing the next packet.  An atom may also contain internal
state that is local to that atom alone and persists across packets. An atom's
body of sequential code fully specifies the atom's behavior and serves as an
interface between the compiler and the programmable switch hardware.

Using this representation, a switch counter that wraps around at a
value of 100 can be written as the atom:\footnote{We use {\tt p.x} to
  represent field {\tt x} within a packet {\tt p} and {\tt x} to
  represent a state variable {\tt x} that persists across packets.}
\begin{lstlisting}[style=customc, numbers=none, frame=none]
if (counter < 99)
  counter++;
else
  counter = 0;
\end{lstlisting}
Similarly, a stateless operation like setting a packet field
(e.g. P4's {\tt modify\_field} primitive~\cite{p4spec}) can be written
as the atom:
\begin{lstlisting}[style=customc, numbers=none, frame=none]
  p.field = value;
\end{lstlisting}
Table~\ref{tab:templates} provides more examples of atoms.

We note that---unlike stateful atomic operations such as a counter---stateless
atomic operations are easier to support with basic packet-field arithmetic.
Consider, for instance, the operation {\tt pkt.f1 = pkt.f2 + pkt.f3 - pkt.f4}.
This operation does not modify any persistent switch state because it only
reads and writes packet fields. It can be implemented without violating
atomicity by using two atoms: one atom to add fields f2 and f3 in one pipeline
stage (clock cycle), and another to subtract f4 from the result in the
next---without having to provide one large atom that supports the entire
operation.

\subsection{Constraining atoms}
\label{s:atomConstraints}

\begin{figure}[h]
  \begin{subfigure}{0.4\columnwidth}
  \begin{center}
  \includegraphics[width=\columnwidth]{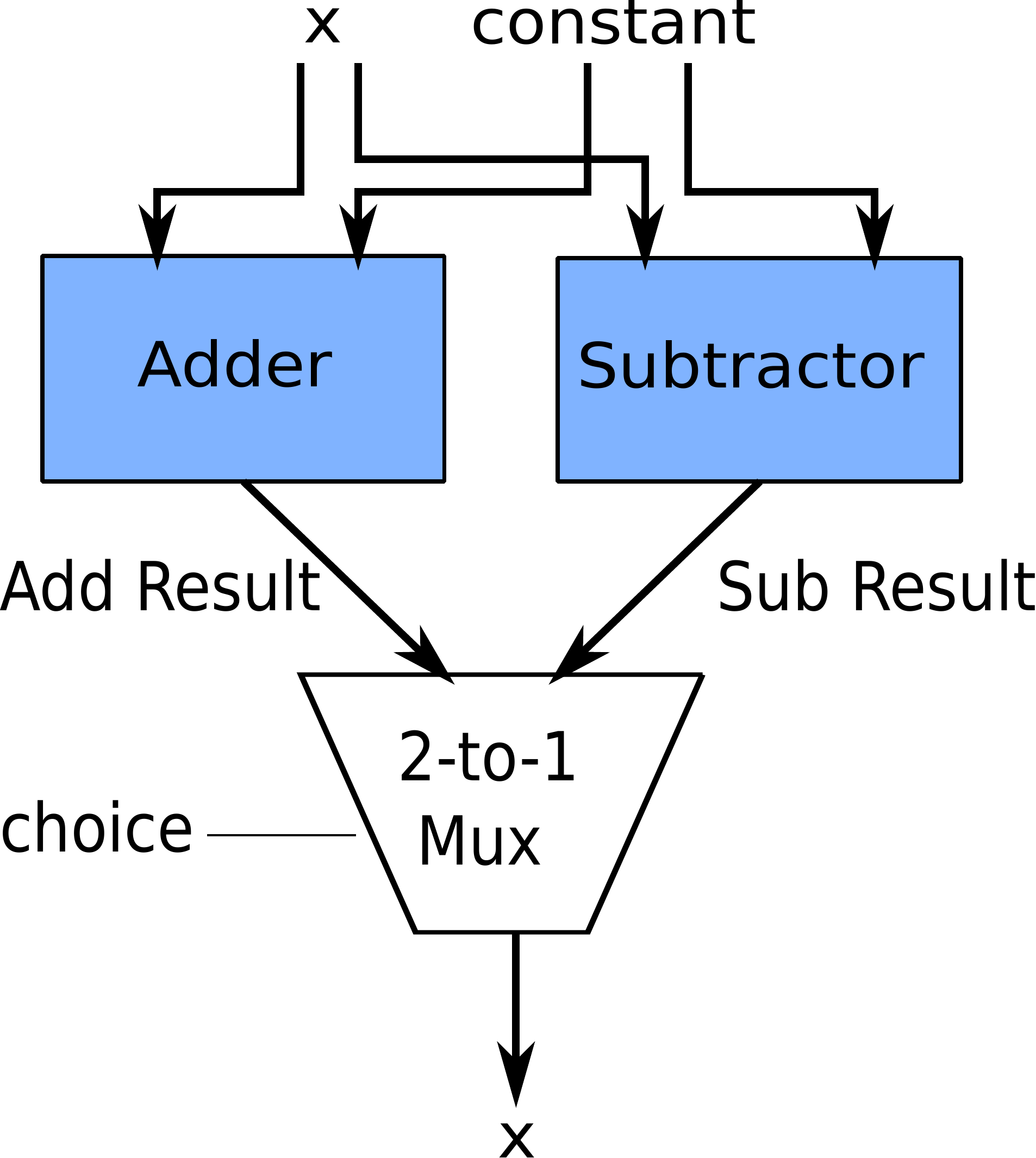}
  \end{center}
  \caption{Circuit for an atom that can add or subtract a constant from a state variable.}
  \label{fig:alu_diag}
  \end{subfigure}
  \hspace{0.05\columnwidth}
  \begin{subfigure}{0.55\columnwidth}
  \begin{lstlisting}
  bit choice = ??;
  int constant = ??;
  if (choice) {
    x = x + constant;
  } else {
    x = x - constant;
  }
  \end{lstlisting}
  \caption{Circuit representation as an atom template.}
  \label{fig:alu_in_sketch}
  \end{subfigure}
  \caption{Atoms and atom templates}
  \label{fig:atom}
\end{figure}

\textbf{Computational limits:} To provide line-rate performance, atom
bodies must finish execution within one clock cycle. We
constrain atom bodies by defining {\it atom templates}
(\S\ref{ss:code_gen}).  An atom template is a program that always
terminates and specifies exactly how the atom is executed. One example
is an ALU with a restricted set of primitive operations to choose from
(Figure~\ref{fig:alu_diag}). Atom templates allow us to create
\absmachine machines with different atoms.  In practice, atom
templates will be designed by an ASIC engineer and exposed as a
machine's instruction set~(\S\ref{ss:targets}).  As programmable
switches evolve, we expect that atoms will evolve as well, but
constrained by the clock-cycle requirement~(\S\ref{ss:perfprog}).

\textbf{Resource limits:} For any real machine, we also need to limit the
number of atoms in each stage (\textit{pipeline width}) and the number of
stages in the pipeline (\textit{pipeline depth}). This is similar to limits on
the number of stages, number of tables per stage, and amount of memory per
stage in programmable switch architectures such as RMT and
FlexPipe~\cite{lavanya_compiler}.

\subsection{What can \absmachine not do?}
\label{ss:limitations}

Like real programmable switches, \absmachine is a good fit for data-plane
algorithms that modify a small set of packet headers and carry out small
amounts of stateful or stateless computation per packet. Data-plane algorithms
like deep packet inspection and WAN optimization require a switch to parse and
process the packet payload as well---effectively parsing a large ``header''
consisting of each byte in the payload, which is challenging at line rates of 1
GHz. Such algorithms are best left to general-purpose CPU platforms~\cite{e2,
aplomb, opennf}. Some algorithms require complex computations, but not on every
packet.  For example, consider a measurement algorithm that periodically scans
a large table to perform garbage collection.  \absmachine's atoms model small
computations that occur on every packet, and are not suitable for such
operations that span many clock cycles.

\section{Packet transactions}
\label{s:transactions}

\begin{figure*}[!t]
\begin{subfigure}{0.5\textwidth}
\begin{small}
\begin{lstlisting}[style=customc]
#define NUM_FLOWLETS    8000
#define THRESHOLD       5
#define NUM_HOPS        10

struct Packet {
  int sport;
  int dport;
  int new_hop;
  int arrival;
  int next_hop;
  int id; // array index
};

int last_time [NUM_FLOWLETS] = {0};
int saved_hop [NUM_FLOWLETS] = {0};

void flowlet(struct Packet pkt) {
  pkt.new_hop = hash3(pkt.sport,
                      pkt.dport,
                      pkt.arrival)
                % NUM_HOPS;

  pkt.id  = hash2(pkt.sport,
                  pkt.dport)
            % NUM_FLOWLETS;

  if (pkt.arrival - last_time[pkt.id] @\label{line:ifStart}@
      > THRESHOLD)
  { saved_hop[pkt.id] = pkt.new_hop; } @\label{line:ifEnd}@

  last_time[pkt.id] = pkt.arrival;
  pkt.next_hop = saved_hop[pkt.id];
}
\end{lstlisting}
\end{small}
\caption{Flowlet switching written in \pktlanguage}
\label{fig:flowlet_code}
\end{subfigure}
\begin{subfigure}{0.4\textwidth}
\includegraphics[width=0.9\columnwidth]{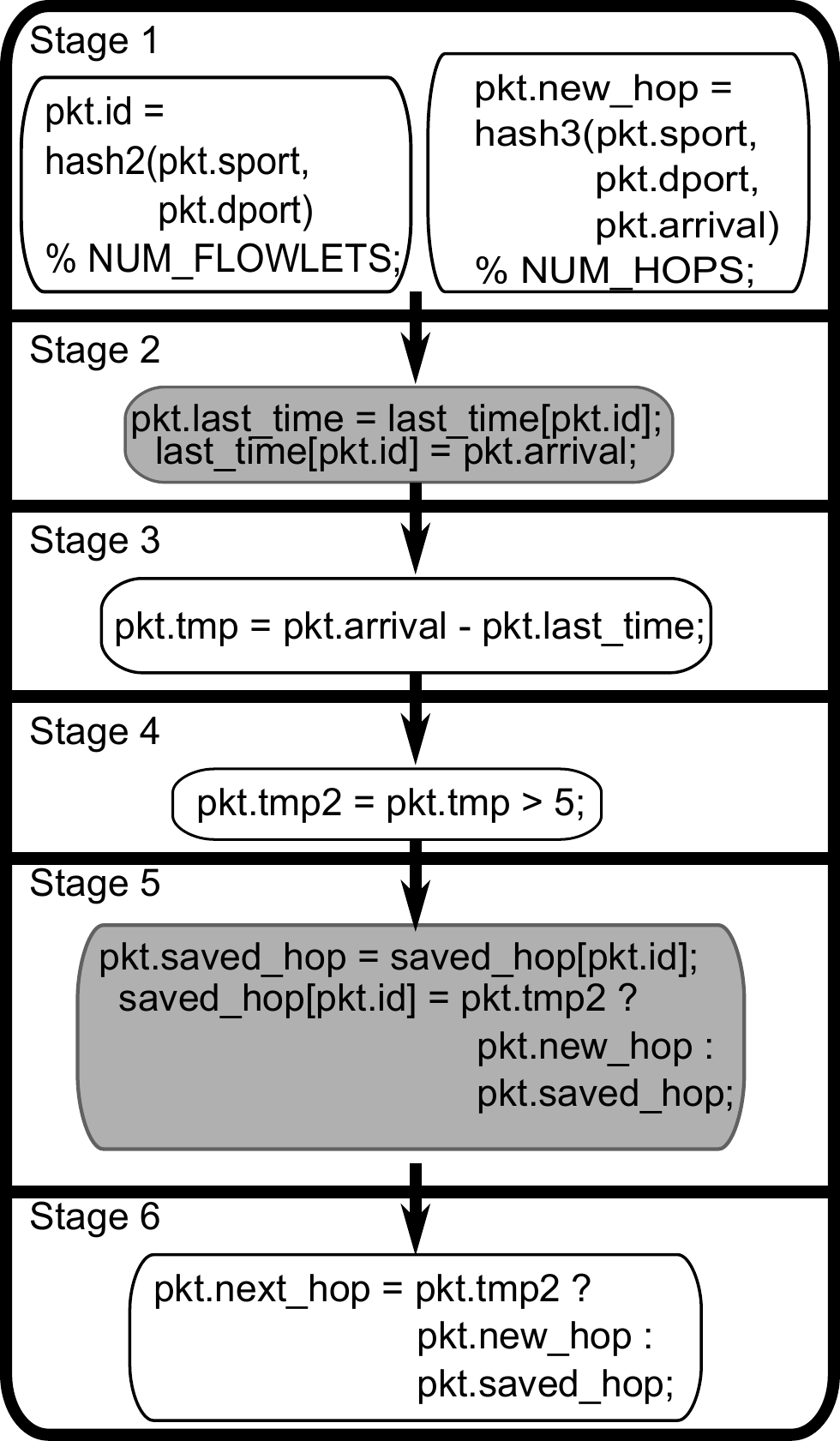}
\caption{6-stage \absmachine pipeline for flowlet
switching.  Control flows from top to bottom. Stateful atoms are in grey.}
\label{fig:flowlet_pipeline}
\end{subfigure}
\caption{Programming flowlet switching in \pktlanguage}
\end{figure*}

To program a data-plane algorithm, a programmer would write code in
\pktlanguage using packet transactions (Figure~\ref{fig:flowlet_code})
and then use the \pktlanguage compiler to compile to an atom pipeline
for a \absmachine machine (Figure~\ref{fig:flowlet_pipeline}). We
first describe packet transactions in greater detail by walking
through an example (\S\ref{ss:flowlet}). Next, we discuss constraints
in \pktlanguage (\S\ref{ss:constraints}) informed by the domain of
line-rate switches. We then discuss triggering packet transactions
(\S\ref{ss:guards}) and handling multiple transactions
(\S\ref{ss:multiple}).

\subsection{\pktlanguage by example}
\label{ss:flowlet}

We use flowlet switching~\cite{flowlets} as an example. Flowlet
switching is a load-balancing algorithm that sends bursts of packets
(called flowlets) from a TCP flow on different paths, provided the
bursts are separated by a large enough time interval to ensure packets
do not arrive out of order at a TCP
receiver. Figure~\ref{fig:flowlet_code} shows flowlet switching in
\pktlanguage. For simplicity, we hash only the source and destination
ports; it is easy to extend it to the full 5-tuple.

This example demonstrates the core language constructs in
\pktlanguage. All packet processing happens in the context of a packet
transaction (the function \texttt{flowlet} starting at line 17). The
function's argument type {\tt Packet} declares the fields in a packet (lines
5--12)\footnote{We use fields to refer to both packet headers such as
  source port ({\tt sport}) and destination port ({\tt dport}) and
  packet metadata ({\tt id}).} that can be referenced by the function
body (lines 18--32).  The function body can also modify persistent
switch state using global variables (e.g.  \texttt{last\_time} and
\texttt{saved\_hop} on lines 14 and 15, respectively).

Conceptually, the switch invokes the packet transaction function one packet at
a time, with no concurrent packet processing. To the programmer, the function
modifies the passed-in packet argument and runs to completion before processing
the next packet.  The function may invoke \textit{intrinsics} such as
\texttt{hash2} on line 23 to use hardware accelerators such as hash generators.
The \pktlanguage compiler uses an intrinsic's signature to infer dependencies
and supplies a canned run-time implementation, but otherwise does not analyze
an intrinsic's internal behavior. When compiled to a \absmachine machine, the
compiler converts the code in Figure~\ref{fig:flowlet_code} to the atom
pipeline in Figure~\ref{fig:flowlet_pipeline}.

\subsection{Constraints on the language}
\label{ss:constraints}

The syntax of Domino is similar to C, but with several constraints
(Table~\ref{tab:restrict}).  These constraints are required for deterministic
performance.  Memory allocation, unbounded iteration counts, and unstructured
control flow all cause variable performance, which may prevent an algorithm
from achieving line rate.  Additionally, \pktlanguage constrains array
modifications by requiring that all accesses to a given array within one
execution of a transaction, i.e. one packet, must use the same array index. For
example, all read and write accesses to the array \texttt{last\_time} use the
index \texttt{pkt.id}, which is constant for each packet, but can change
between packets. This restriction mirrors restrictions on memories, which don't
typically support distinct read and write addresses every clock cycle.

\begin{table}
  \begin{tabular}{p{0.9\columnwidth}}
    No iteration (while, for, do-while).\\
    No goto, break, or continue.\\
    No pointers.\\
    No dynamic memory allocation / heap.\\
    Array index is constant for each transaction execution.\\
    No access to data i.e. unparsed portion of the packet.\\
  \end{tabular}

  \caption{Restrictions in \pktlanguage}

  \label{tab:restrict}
\end{table}

\subsection{Triggering packet transactions}
\label{ss:guards}
Packet transactions specify \textit{how} to process packet headers and/or
state.  To specify {\em when} to run packet transactions, we provide a {\em
guard}: a predicate on packet fields that triggers the transaction whenever a
packet matches the guard. An example guard (pkt.tcp\_dst\_port == 80) would execute
heavy-hitter detection on all packets on TCP destination port 80. This guard can
be implemented using an exact match in a match-action table, with the actions
being the atoms resulting from compiling the packet transaction. Guards can be of various forms,
e.g., exact, ternary, longest-prefix and range-based matches, depending on the
match semantics supported by the match-action pipeline. Because guards map
rather straightforwardly to the match key in a match-action table, this paper only
focuses on compiling packet transactions.

\subsection{Handling multiple transactions}
\label{ss:multiple}
So far, we have discussed a single packet transaction corresponding to a single
data-plane algorithm. In practice, a switch
would run multiple data-plane algorithms---each processing its own subset of
packets. To accommodate multiple transactions, we envision a policy
language that specifies pairs of guards and transactions. Realizing a policy is
straightforward when all guards are disjoint. When guards overlap, multiple
transactions need to execute on the same subset of packets, requiring a
mechanism to compose transactions. One semantics for composition is to concatenate the two
transaction bodies in an order specified by the user, providing the illusion of
a larger transaction that combines two transactions. We leave a detailed
exploration of this and alternative semantics to future work,
and focus only on compiling a single packet transaction.

\section{The Domino compiler}
\label{s:compiler}

The \pktlanguage compiler compiles from \pktlanguage programs to \absmachine
targets. The compiler provides an {\em all-or-nothing model}: if compilation
succeeds, the compiler guarantees that the program will run at line rate on the
target. If the program can't be run at line rate, the compiler rejects the
program outright; there is no smooth tradeoff between a program's performance
and its complexity.  This all-or-nothing compilation model is unusual relative
to other substrates such as a CPU, GPU, or DSP. But, it reflects how routers
are used today. Routers are rated for a particular line rate, regardless of the
enabled feature set. The all-or-nothing model trades off diminished
programmability for guaranteed line-rate performance, in contrast to software
routers that provide greater flexibility but unpredictable run-time
performance~\cite{dobrescu2012, wenfei15}.

\begin{figure}[!t]
  \includegraphics[width=\columnwidth]{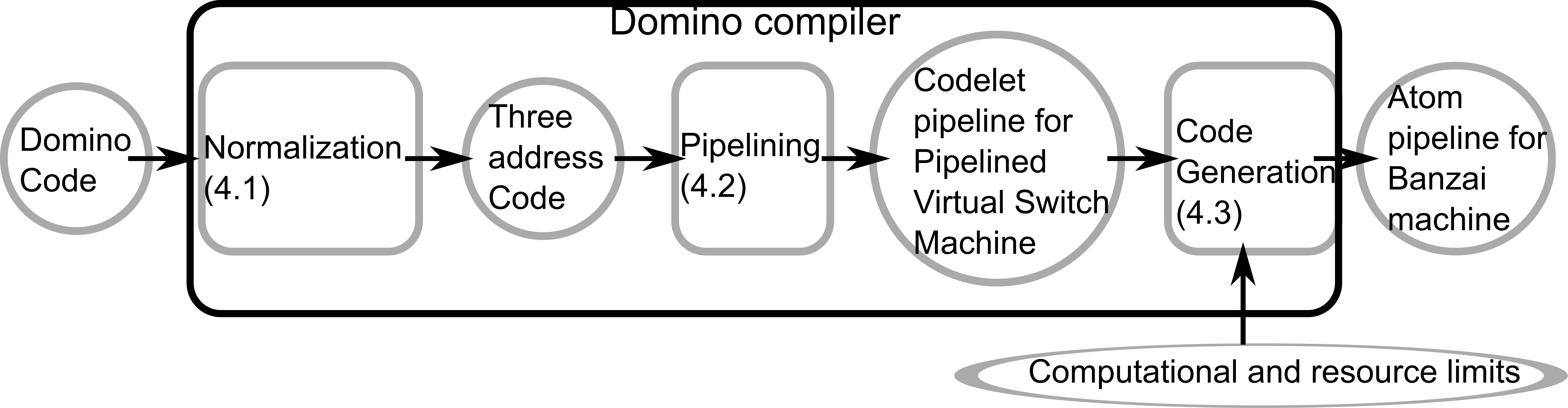}
  \caption{Passes in the \pktlanguage compiler}
  \label{fig:passes}
\end{figure}

The \pktlanguage compiler has three passes (Figure~\ref{fig:passes}).  First,
\textit{normalization} simplifies the packet transaction into a restrictive
three-address code form while retaining the sequential nature of packet
transactions, i.e., processing one packet at a time. Second, \textit{pipelining} transforms the normalized code into
code for a \textit{pipelined virtual switch machine (PVSM)}. PVSM is an
intermediate representation that models a switch pipeline with no computational
or resource limits. Third, \textit{code generation} transforms this
intermediate representation into configuration for a \absmachine machine, given
as inputs the machine's computational and resource constraints, and rejects the
program if it can't run at line rate on that \absmachine machine.  The
\pktlanguage compiler uses many existing compiler techniques, but adapts and
simplifies them in important ways to suit the domain of line-rate switches
(\S\ref{ss:related_compiler}). Throughout this section, we use flowlet
switching as a running example to demonstrate compiler passes.

\subsection{Normalization}
\label{ss:normalization}

\textbf{Branch removal: }A packet transaction's body can contain (potentially
nested) branches (e.g., Lines~\ref{line:ifStart} to \ref{line:ifEnd} in
Figure~\ref{fig:flowlet_code}).  Branches alter control flow and complicate
dependency analysis, i.e.,  whether a statement should precede another.  We
transform branches into the conditional operator, starting from the innermost
\texttt{if} and recursing outwards (Figure~\ref{fig:if_convert}).  This turns
the transaction body into straight-line code with no branches.  Straight-line
code simplifies the rest of the compiler, by simplifying dependency analysis
and conversion to static single-assignment form.

\textbf{Rewriting state variable operations: }We now identify state variables
in a packet transaction, such as \texttt{last\_time} and \texttt{saved\_hop} in
Figure~\ref{fig:flowlet_code}.  For each state variable, we create a
\textit{read flank} to read the state variable into a temporary packet field.
For an array, we also move the index expression into the read flank using the
fact that only one array index is accessed by each packet.  Within the packet
transaction, we replace the state variable with the packet temporary, and
create a \textit{write flank} to write the packet temporary back into the state
variable~(Figure~\ref{fig:stateful_flanks}). After this, the only operations
on state variables are reads and writes; all arithmetic happens on packet
fields. Restricting stateful operations simplifies handling of state during
pipelining.

\textbf{Converting to static single-assignment form: }We next convert the code to
static single-assignment form (SSA)~\cite{ssa}, where every packet field is
assigned exactly once. To do so, we replace every assignment to a packet field
with a new packet field and propagate this until the next assignment to the
same field~(Figure~\ref{fig:ssa}) .  Because every field is assigned exactly
once, SSA removes Write-After-Read and Write-After-Write dependencies.  Only
Read-After-Write dependencies remain, simplifying dependency analysis.

\textbf{Flattening to three-address code: } Three-address code~\cite{tac} is a
representation where all instructions are either reads/writes into state
variables or operations on packet fields of the form \texttt{pkt.f1 = pkt.f2 op
pkt.f3;} where \texttt{op} can be an arithmetic, logical, relational, or conditional
\footnote{Conditional operations alone have 4 arguments.} operator.  We also allow either one of {\tt pkt.f2} or {\tt pkt.f3}
to be an intrinsic function call.  To convert to three-address code, we flatten
expressions that are not in three-address code using
temporaries~(Figure~\ref{fig:three_address}).

\subsection{Pipelining}
\label{ss:partitioning}
At this point, the normalized code is still sequential in that it operates on a
single packet at a time without using a pipeline to process packets
concurrently.  Pipelining turns sequential code into a pipeline of
\textit{codelets}, where each codelet is a sequential block of three-address
code statements. This codelet pipeline corresponds to an intermediate
representation (IR) we call the \textit{Pipelined Virtual Switch Machine
(PVSM)}. PVSM places no computational or resource constraints on the
pipeline---much like IRs such as LLVM place no restriction on the number of
virtual registers. Later, during code generation, we map these codelets to
atoms available in a \absmachine machine, while respecting its constraints.

\begin{figure*}[!t]
\begin{subfigure}{0.5\textwidth}
  \includegraphics[width=0.8\columnwidth]{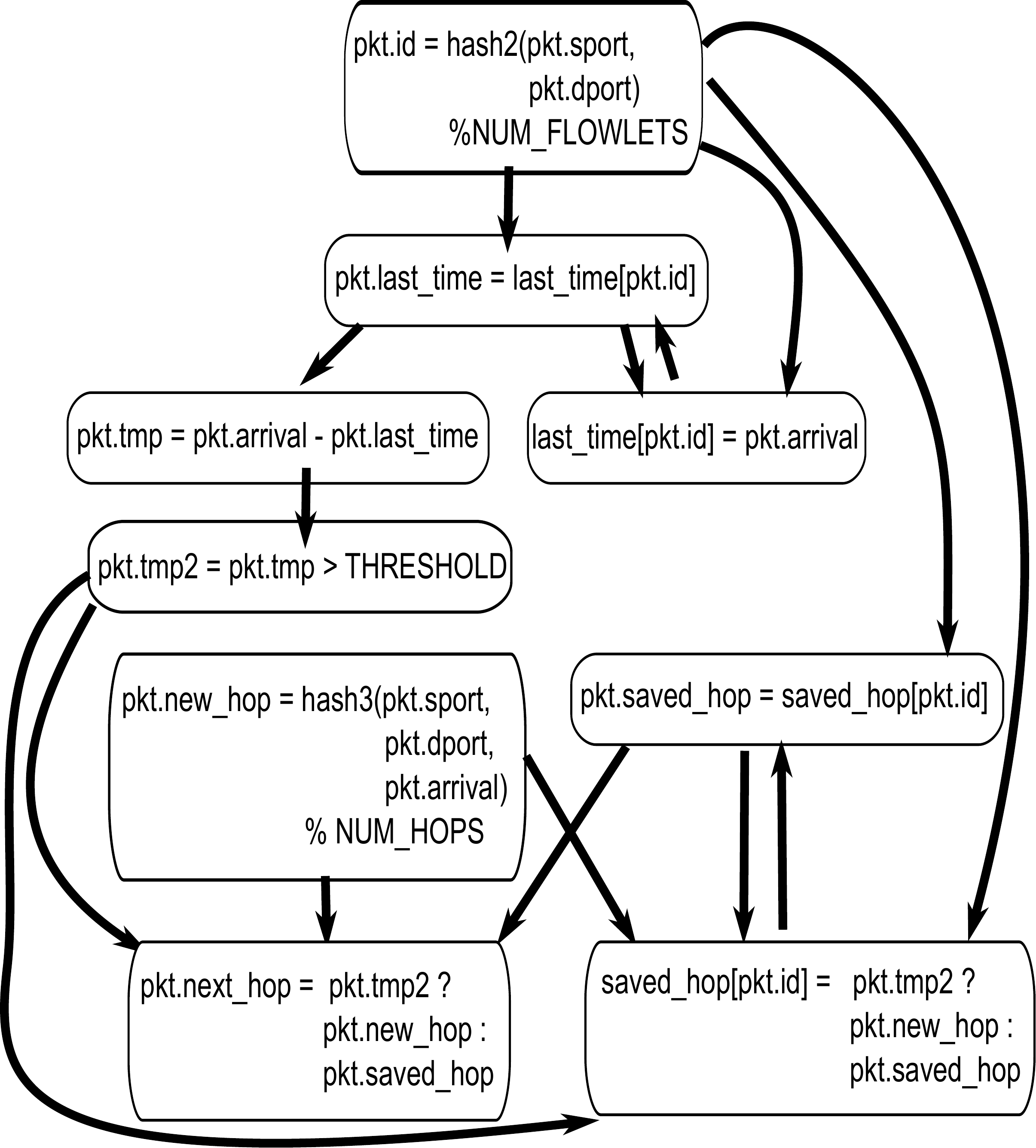}
  \caption{Dependency graph. Edges are read-after-write dependencies.}
  \label{fig:partitioning_before}
\end{subfigure}
$\Longrightarrow$ 
\begin{subfigure}{0.5\textwidth}
\includegraphics[width=0.8\columnwidth]{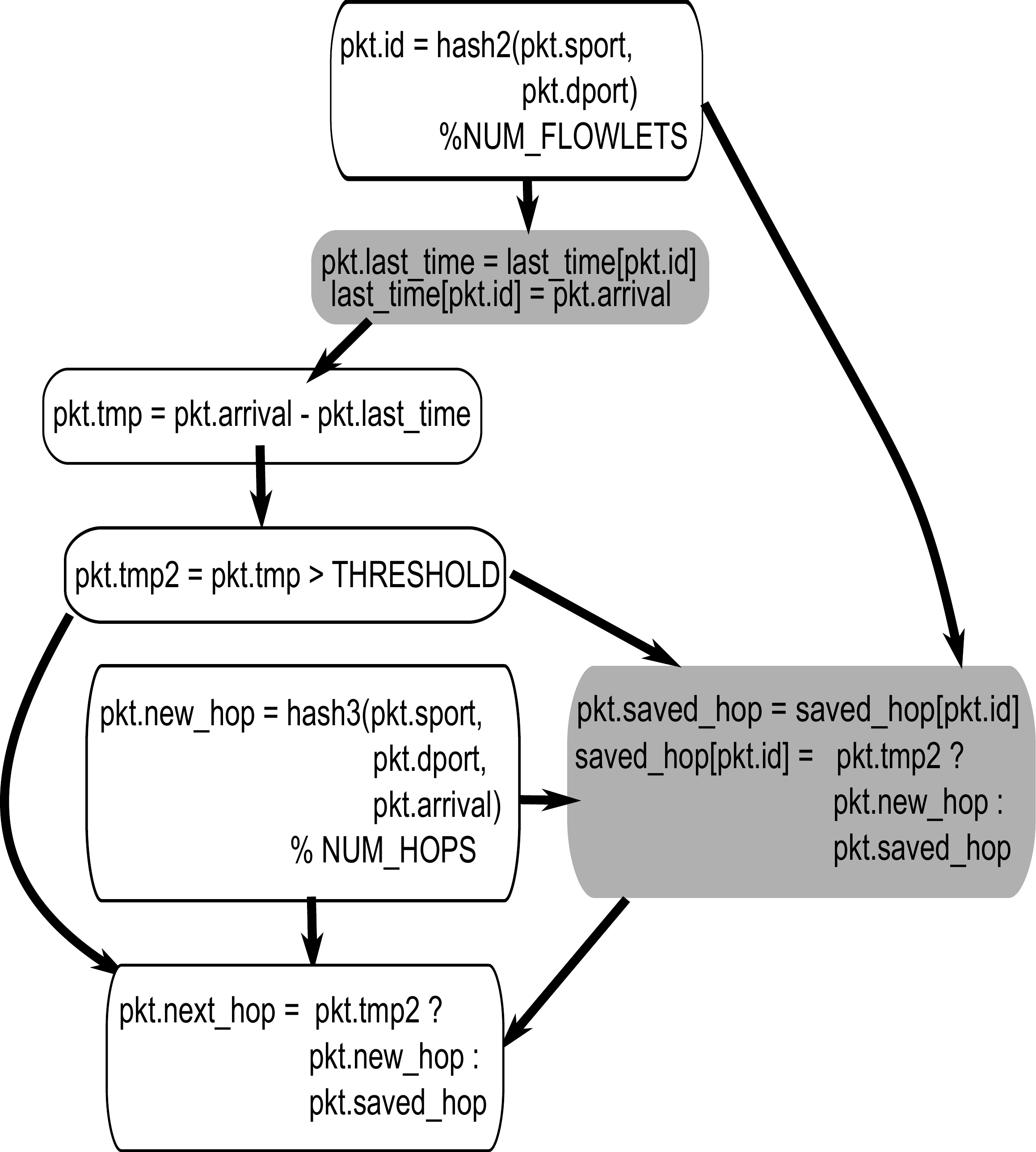}
\caption{DAG after condensing SCCs.}
\label{fig:partitioning_after}
\end{subfigure}
\caption{Code Pipelining}
\label{fig:pipelining}
\end{figure*}

We create PVSM's codelet pipeline using the steps below.
\begin{CompactEnumerate}
  \item Create a dependency graph of all statements in the normalized packet
    transaction. First, create a node for each statement. Second,
    add a pair of edges between any two nodes N1 and N2, where N1 is a read
    from a state variable and N2 is a write into the same variable, to capture
    the notion that state should be internal to a codelet/atom. Third, create
    an edge (N1, N2) for every pair of nodes N1, N2 where N2 reads a variable
    written by N1.  We only check read-after-write dependencies because we
    eliminate control dependencies\footnote{An instruction A is control
    dependent on a preceding instruction B if the outcome of B determines
    whether A should be executed or not.} through branch removal, and
    write-after-read and write-after-write dependencies don't exist after SSA.
    Figure~\ref{fig:partitioning_before} shows the resulting dependency graph.
  \item Generate strongly connected components (SCCs) of this dependency graph
    and condense them into a directed acyclic graph (DAG). This captures the notion that all
    operations on a state variable must be confined to one codelet/atom because
    state cannot be shared between atoms. Figure~\ref{fig:partitioning_after}
    shows the resulting DAG.
  \item Schedule the resulting DAG using critical path
    scheduling~\cite{crit_path_sched} by creating a new pipeline stage when one
    operation needs to follow another. This results in the codelet pipeline
    shown in Figure~\ref{fig:flowlet_pipeline}.\footnote{We refer to this both
    as a codelet and an atom pipeline because codelets map one-to-one atoms
  (\S\ref{ss:code_gen}).}
\end{CompactEnumerate}

The codelet pipeline implements the packet transaction on a switch pipeline
with no computational or resource constraints. We handle these constraints
next.

\subsection{Code generation}
\label{ss:code_gen}

To determine if the codelet pipeline can be compiled to a \absmachine machine,
we consider two constraints in any \absmachine machine: resource limits, i.e.,
the pipeline width and depth, and computational limits on atoms within a
pipeline stage, i.e., the atom templates provided by a \absmachine machine.

\textbf{Resource limits:} To handle resource limits, we scan each pipeline
stage in the codelet pipeline starting from the first to check for pipeline
width violations.  If we violate the pipeline width, we insert as many new
stages as required and spread codelets evenly across these stages.  We continue
until the number of codelets in all stages is under the pipeline width and
reject the program if we exceed the pipeline depth.

\textbf{Computational limits:} Next, we determine if codelets in the pipeline
map one-to-one to atoms provided by the \absmachine machine. In general,
codelets have multiple three-address code statements that need to execute
atomically. For instance, updating the state variable \texttt{saved\_hop} in
Figure~\ref{fig:flowlet_pipeline} requires a read followed by a conditional
write.  It is not apparent whether such codelets can be mapped to an available
atom. We develop a new technique to determine the implementability of a codelet,
given an atom template.

Each atom template has a set of configuration parameters, where the parameters
determine the atom's behavior.  For instance, Figure~\ref{fig:alu_diag} shows a
hardware circuit that can perform stateful addition or subtraction, depending
on the value of the constant and which output is selected from the multiplexer.
Its atom template is shown in Figure~\ref{fig:alu_in_sketch}, where {\tt
choice} and {\tt constant} represent configuration parameters.  Each codelet
can be viewed as a functional specification of the atom.  With that in mind,
the mapping problem is equivalent to searching for the value of the parameters
to configure the atom such that it implements the provided specification.

We use the SKETCH program synthesizer~\cite{sketch_asplos} for this purpose, as
the atom templates can be easily expressed using SKETCH, while SKETCH also
provides efficient search algorithms and has been used for similar purposes in
other domains~\cite{bitstreaming, lifejoin, qbs, chlorophyll}.  As an
illustration, assume we want to map the codelet {\tt x=x+1} to the atom
template shown in Figure~\ref{fig:alu_in_sketch}. SKETCH will search for
possible parameter values so that the resulting atom is functionally identical
to the codelet, for all possible input values of {\tt x}.  In this case, SKETCH
finds the solution with {\tt choice=0} and {\tt constant=1}.  In contrast, if
the codelet {\tt x=x*x} was supplied as the specification, SKETCH will return
an error as no parameters exist.

\begin{figure*}[!t]
  \hspace{-0.3in}
  \begin{minipage}{0.55\textwidth}
  \begin{small}
  \begin{lstlisting}[style=customc, numbers=none, frame=none]
  if (@\textcolor{blue}{pkt.arrival - last\_time[pkt.id] > THRESHOLD}@) {
    saved_hop[pkt.id] = pkt.new_hop;
  }
  \end{lstlisting}
  \end{small}
  \end{minipage}
  \hspace{-0.5in}
  $\Longrightarrow$ 
  \hspace{-0.3in}
  \begin{minipage}{0.6\textwidth}
  \begin{small}
  \begin{lstlisting}[style=customc, numbers=none, frame=none]
  @\textcolor{blue}{pkt.tmp = pkt.arrival - last\_time[pkt.id]  > THRESHOLD}@;
  saved_hop[pkt.id] = @\textcolor{blue}{pkt.tmp}@
                      ? pkt.new_hop
                      : saved_hop[pkt.id]; @\textcolor{magenta}{// Rewritten}@
  \end{lstlisting}
  \end{small}
  \end{minipage}
%\vspace{-.2in}
\caption{Branch removal}
\label{fig:if_convert}
\end{figure*}

\begin{figure*}[!t]
  \begin{minipage}{0.43\textwidth}
  \begin{small}
  \begin{lstlisting}[style=customc, numbers=none, frame=none]
pkt.id = hash2(pkt.sport,
               pkt.dport)
         % NUM_FLOWLETS;
...
@\textcolor{blue}{last\_time[pkt.id] = pkt.arrival;}@
...
  \end{lstlisting}
  \end{small}
  \end{minipage}
  \hspace{-0.5in}
  $\Longrightarrow$ 
  \hspace{-0.2in}
  \begin{minipage}{0.61\textwidth}
  \begin{small}
  \begin{lstlisting}[style=customc, numbers=none, frame=none]
pkt.id = hash2(pkt.sport,           @\textcolor{magenta}{// Read flank}@
               pkt.dport)
         % NUM_FLOWLETS;
pkt.last_time = last_time[pkt.id];  @\textcolor{magenta}{// Read flank}@
...
@\textcolor{blue}{pkt.last\_time = pkt.arrival;}@             @\textcolor{magenta}{// Rewritten}@
...
last_time[pkt.id] = pkt.last_time;  @\textcolor{magenta}{// Write flank}
  \end{lstlisting}
  \end{small}
  \end{minipage}
  \caption{Rewriting state variable operations}
\label{fig:stateful_flanks}
\end{figure*}

\begin{figure*}[!t]
  \begin{minipage}{\textwidth}
  \begin{minipage}{0.4\textwidth}
  \begin{small}
  \begin{lstlisting}[style=customc, numbers=none, frame=none]
@\textcolor{blue}{pkt.id}@ = hash2(pkt.sport,
              pkt.dport)
              % NUM_FLOWLETS;
@\textcolor{blue}{pkt.last\_time}@ = last_time[@\textcolor{blue}{pkt.id}@];
...
@\textcolor{blue}{pkt.last\_time}@ = pkt.arrival;
last_time[@\textcolor{blue}{pkt.id}@] = @\textcolor{blue}{pkt.last\_time}@;
  \end{lstlisting}
  \end{small}
  \end{minipage}
 % 
  %\hspace{-0.1in}
  $\Longrightarrow$
  \hspace{-0.2in}
  \begin{minipage}{0.6\textwidth}
  \begin{small}
  \begin{lstlisting}[style=customc, numbers=none, frame=none]
@\textcolor{blue}{pkt.id0}@ = hash2(pkt.sport,          @\textcolor{magenta}{// Rewritten}@ @\label{line:assign}@
               pkt.dport)
               % NUM_FLOWLETS;  
@\textcolor{blue}{pkt.last\_time0}@ = last_time[@\textcolor{blue}{pkt.id0}@];  @\textcolor{magenta}{// Rewritten}@
...
@\textcolor{blue}{pkt.last\_time1}@ = pkt.arrival;        @\textcolor{magenta}{// Rewritten}@
last_time[@\textcolor{blue}{pkt.id0}@] = @\textcolor{blue}{pkt.last\_time1}@;  @\textcolor{magenta}{// Rewritten}@
  \end{lstlisting}
  \end{small}
  \end{minipage}
  \caption[title]{Converting to static single-assignment form}
  \label{fig:ssa}
\end{minipage}
\end{figure*}

\begin{figure*}[!t]
\begin{minipage}{\textwidth}
\begin{lstlisting}[style=customc]
pkt.id            = hash2(pkt.sport, pkt.dport) % NUM_FLOWLETS; @\label{line:id}@
pkt.saved_hop     = saved_hop[pkt.id]; @\label{line:stateRead}@
pkt.last_time     = last_time[pkt.id];
pkt.new_hop       = hash3(pkt.sport, pkt.dport, pkt.arrival) % NUM_HOPS; @\label{line:newhop}@
pkt.tmp           = pkt.arrival - pkt.last_time;
pkt.tmp2          = pkt.tmp > THRESHOLD;
pkt.next_hop      = pkt.tmp2 ? pkt.new_hop : pkt.saved_hop;
saved_hop[pkt.id] = pkt.tmp2 ? pkt.new_hop : pkt.saved_hop; @\label{line:stateWrite}@
last_time[pkt.id] = pkt.arrival;
\end{lstlisting}
\caption[title2]{Flowlet switching in three-address
code. Lines~\ref{line:id} and \ref{line:newhop} are flipped relative
to Figure~\ref{fig:flowlet_code} because {\tt pkt.id} is an array index expression and is
moved into the read flank.}
\label{fig:three_address}
\end{minipage}
\vspace{-0.3cm}
\end{figure*}

\subsection{Related compiler techniques}
\label{ss:related_compiler}

The \pktlanguage compiler employs many techniques from the compiler literature,
but adapts and simplifies them in new ways to suit the domain of line-rate
switches (Table~\ref{tab:prior_compiler}). The use of SCCs is inspired by
software pipelining for VLIW architectures~\cite{software_pipelining, rau}. The size
of the largest SCC affects the {\em maximum throughput} of the pipelined loop
in software pipelining. For \pktlanguage, it affects the {\em circuit area} of
the atom required to run a program at line rate. \pktlanguage trades off an
increase in space for line-rate performance.

Program synthesis was used for code generation in
Chlorophyll~\cite{chlorophyll}.  Code generation for \pktlanguage shares
similar goals as technology mapping~\cite{micheli, flowmap, spectransform} and
instruction selection~\cite{muchnik}.  However, prior work maps a code sequence
to \textit{multiple} instructions/tiles, using heuristics to minimize
instruction count. Domino's problem is simpler: we map each codelet to a single
atom using SKETCH.  The simpler problem allows a non-heuristic solution: if
there is any way to map the codelet to an atom, SKETCH will find it.

Branch removal resembles if-conversion~\cite{if_conversion}, a
technique used in vectorizing compilers. This procedure is easier in Domino
because there is no backward control transfer (goto, break,
continue). Domino's SSA computation operates on straight-line code and doesn't
 handle branches, which considerably complicate SSA algorithms~\cite{ssa}.

\begin{table}[!t]
  \begin{scriptsize}
    \begin{tabular}{|p{0.1\textwidth}|p{0.1\textwidth}|p{0.2\textwidth}|}
  \hline
  Technique & Prior Work & Differences \\
  \hline
  Conversion to straight-line code & If Conversion~\cite{if_conversion} & No backward control flow (gotos, break, continue) \\
  \hline
  SSA & Cytron et. al~\cite{ssa} & SSA runs on straight-line code with no branches \\
  \hline
  Strongly Connected Components & Lam~\cite{software_pipelining}, Rau and Glaeser~\cite{rau} & Scheduling in space vs. time \\
  \hline
  Code generation using program synthesis & Chlorophyll~\cite{chlorophyll}, technology mapping~\cite{micheli, flowmap, spectransform}, instruction selection~\cite{muchnik} & Optimal vs. best-effort mapping; One-to-one mapping vs. tiling \\
  \hline
  \end{tabular}
  \end{scriptsize}
  \caption{Domino's compiler in relation to prior work}
  \label{tab:prior_compiler}
\end{table}

\section{Evaluation}
\label{s:eval}

\begin{table}[!t]
  \begin{scriptsize}
  \begin{tabular}{|p{0.13\textwidth}|p{0.22\textwidth}|p{0.04\textwidth}|}
    \hline
    Atom & Description & Area (\si{\micro\metre\squared})\\
    \hline
    Stateless & Arithmetic, logic, relational, and conditional operations on packet/constant operands & 1384 \\
    \hline
    Read/Write & Read/Write packet field/constant into single state variable. & 250 \\
    \hline
    ReadAddWrite (RAW) & Add packet field/constant to state variable (OR) Write packet field/constant into state variable. & 431 \\
    \hline
    Predicated ReadAddWrite (PRAW) & Execute RAW on state variable only if a predicate is true, else leave unchanged. & 791 \\
    \hline
    IfElse ReadAddWrite (IfElseRAW) & Two separate RAWs: one each for when a predicate is true or false. & 985 \\
    \hline
    Subtract (Sub) & Same as IfElseRAW, but also allow subtracting a packet field/constant. & 1522 \\
    \hline
    Nested Ifs (Nested) & Same as Sub, but with an additional level of nesting that provides 4-way predication. & 3597 \\
    \hline
    Paired updates (Pairs) & Same as Nested, but allow updates to a pair of state variables, where predicates can use both state variables. & 5997 \\
    \hline
  \end{tabular}
  \end{scriptsize}
  \caption{Atom areas in a 32 nm standard-cell library.  All atoms meet timing
  at 1GHz. Each of the seven compiler targets contains one of the seven
  stateful atoms (Read/Write through Pairs) and the single stateless atom.}
  \label{tab:templates}
\end{table}

\begin{table*}[!t]
  \begin{tabular}{|p{0.16\textwidth}|p{0.34\textwidth}|p{0.08\textwidth}|p{0.09\textwidth}|p{0.07\textwidth}|p{0.06\textwidth}|p{0.05\textwidth}|}
\hline
Algorithm & Description & Least expressive atom & \# of stages, max. atoms/stage & Ingress or Egress Pipeline? & Domino LOC & P4 LOC\\
\hline
Bloom filter~\cite{bloom} (3 hash functions) & Set membership bit on every packet. & Write & 4, 3 & Either & 29 & 104 \\
\hline
Heavy Hitters~\cite{opensketch} (3 hash functions) & Increment Count-Min Sketch~\cite{cormode} on every packet. & RAW & 10, 9 & Either & 35 & 192 \\
\hline
Flowlets~\cite{flowlets} & Update saved next hop if flowlet threshold is exceeded. & PRAW & 6, 2 & Ingress & 37 & 107 \\
\hline
RCP~\cite{rcp} & Accumulate RTT sum if  RTT is under maximum allowable RTT. & PRAW & 3, 3 & Egress & 23 & 75 \\
\hline
Sampled NetFlow~\cite{sampled_nflow} & Sample a packet if packet count reaches N; Reset count to 0 when it reaches N. & IfElseRAW & 4, 2 & Either  & 18 & 70 \\
\hline
HULL~\cite{hull} & Update counter for virtual queue. & Sub & 7, 1 & Egress & 26 & 95 \\
\hline
Adaptive Virtual Queue~\cite{avq} & Update virtual queue size and virtual capacity & Nested & 7, 3 & Ingress & 36 & 147 \\
\hline
Compute priorities for weighted fair queueing~\cite{pifo_hotnets} & Compute packet's virtual start time using finish time of last packet in that flow. & Nested & 4, 2 & Ingress & 29 & 87 \\
\hline
DNS TTL change tracking~\cite{dns_change} & Track number of changes in announced TTL for each domain & Nested & 6,3 & Ingress & 27 & 119 \\
\hline
CONGA~\cite{conga} & Update best path's utilization/id if we see a better path. Update best path utilization alone if it changes. & Pairs & 4, 2 & Ingress & 32 & 89\\
\hline
CoDel~\cite{codel} & Update: Whether we are marking or not, Time for next mark, Number of marks so far, Time at which min. queueing delay will exceed target. & Doesn't map & 15, 3 & Egress & 57 & 271\\
\hline
\end{tabular}
\caption{Data-plane algorithms}
\label{tab:algos}
\end{table*}

We have performed a number of experiments to evaluate \pktlanguage. 
First, we evaluate \pktlanguage's expressiveness by using it to program several
data-plane algorithms (Table~\ref{tab:algos}), and comparing it to writing them
in P4~(\S\ref{ss:expressiveness}).  To validate that these algorithms can be
implemented at line rate, we design a concrete set of \absmachine machines that
we use as compiler targets for \pktlanguage~(\S\ref{ss:targets}).  We estimate
that these machines are feasible in hardware today because their atoms incur
modest chip area overhead. Next, we use the \pktlanguage compiler to compile
the algorithms in Table~\ref{tab:algos} to these targets~(\S\ref{ss:compiler}).
We conclude by quantifying the tradeoff between a target's programmability (the
space of data-plane algorithms that it can run at line rate) and the target's
performance (the maximum line rate it can support)~(\S\ref{ss:perfprog}).

\subsection{Expressiveness}
\label{ss:expressiveness}

To evaluate \pktlanguage's expressiveness, we express several data-plane
algorithms (Table~\ref{tab:algos}) using \pktlanguage. These algorithms
encompass a variety of data-plane functionality including data-plane load
balancing, in-network congestion control, active queue management, security,
and measurement. In addition, we also used Domino to express the priority
computation for programming scheduling using the push-in first-out queue
abstraction~\cite{pifo_hotnets}. In all these cases, the algorithms are
already available as blocks of imperative code from online sources; 
translating them to \pktlanguage syntax
was straightforward.

In contrast, expressing any of these algorithms in P4 requires manually teasing
out portions of the algorithm that can reside in independent match-action
tables and then chaining these tables together. In essence, the programmer
manually carries out the transformations in \pktlanguage's compiler. Of the
algorithms in Table~\ref{tab:algos}, only flowlet switching has a publicly
available P4 implementation~\cite{p4_flowlet} that we can compare against. This
implementation requires 231 lines of uncommented P4, in comparison to the 37
lines of \pktlanguage code in Figure~\ref{fig:flowlet_code}. Not only that,
using P4 also requires the programmer to manually specify tables, the actions
within the tables, how tables are chained, and what headers are required---all
to implement a single data-plane algorithm. As the \pktlanguage compiler shows,
this process can be automated; to demonstrate this, we developed a backend for
\pktlanguage that generates the equivalent P4 code (lines of code for these auto-generated P4
programs are listed in Table~\ref{tab:algos}).

Lastly, data-plane algorithms on software platforms today (NPUs,
Click~\cite{click}, the Linux qdisc subsystem~\cite{qdisc})  are programmed in languages
resembling \pktlanguage---hence we are confident that the \pktlanguage
syntax is already familiar to network operators.

\subsection{Compiler targets}
\label{ss:targets}

We design a concrete set of compiler targets for \pktlanguage based on the
\absmachine machine model. First, we specify computational limits on atoms in
each compiler target using atom templates. Using the Synopsys Design
Compiler~\cite{synopsys_dc}, we quantify each atom's area in a 32 nm
standard-cell library when running at 1 GHz.  Second, using an individual
atom's area and a switching chip's area~\cite{gibb_parsing}, we determine the
machine's resource limits, i.e., the pipeline width for each atom and the
pipeline depth.

\textbf{Computational limits:}
Stateless atoms are easier to design because arbitrary stateless operations can
be spread out across multiple pipeline stages without violating
atomicity~(\S\ref{ss:atoms}). We design a stateless atom that can support
simple arithmetic (add, subtract, left shift, right shift), logical (and, or,
xor), relational ({\tt >=}, {\tt <=}, {\tt ==}, {\tt !=}), or conditional operations (C's ``{\tt ?}''
operator) on a set of packet fields. Any packet field can also be substituted
with a constant operand.

Designing stateful atoms is more involved because it determines which
algorithms the switch can support. A more complex stateful atom can support
more data-plane algorithms, but occupies greater chip area. To illustrate this,
we design a containment hierarchy of stateful atoms, where each atom can
express all stateful operations that its predecessor can. When synthesized to a
32 nm standard-cell library, all of our designed atoms meet timing at 1 GHz and their area
increases with the atom's complexity~(Table~\ref{tab:templates}).

\textbf{Resource limits:}
We design one compiler target for each combination of a stateful atom along
with the single stateless atom in Table~\ref{tab:templates}.  We determine
resource limits for stateful and stateless atoms separately.  For the stateless
atom, assuming a chip area of 200 \si{\milli\metre\squared} (the smallest area given by Gibb et
al.~\cite{gibb_parsing}), and an acceptable overhead of 7\% (the area overheads
for actions in RMT~\cite{rmt}), we can support \textasciitilde10000 stateless atoms, given
the area of 1384 \si{\micro\metre\squared} per instance.  If these 10000 atoms were
spread across the same number of stages (32) as RMT, we could support up to
\textasciitilde300 stateless atoms per stage.

A similar analysis for the stateful atoms yields 70 stateful atoms per stage for
the most complex stateful atom (Pairs) with an area of 5997 \si{\micro\metre\squared}.
However, stateful atoms access per-stage memory banks storing state.
Providing 70 independent memory banks per stage supporting one read and
write per clock is prohibitive. Furthermore, given an overall stage memory budget,
slicing it into many small banks reduces the amount of memory accessible to
each atom. This is problematic for hash-based algorithms that need to hash into
a large memory space. Taking these into account, we limit the number of
stateful atoms to around 10 per stage, which is still sufficient for the
data-plane algorithms that we are interested in. The area overhead of these 10
stateful atoms is \textasciitilde1\%.

We next look at the multiplexers to route inputs to these atoms from specific
packet fields and route outputs from these atoms to specific packet fields. For
this, we rely on RMT~\cite{rmt}, which estimates a crossbar area of 6 \si{\milli\metre\squared}
for a 32-stage pipeline with 224 action units. Scaling this proportionally to
300 atoms, we estimate a crossbar area of 8 \si{\milli\metre\squared} with a 4\% area
overhead.

In summary, we assume 32 stages in total, 300 stateless atoms per stage and 10
stateful atoms per stage for all compiler targets with an area overhead of 12\%
(7\% for stateless atoms, 1\% for the stateful atoms, and 4\% for the
crossbars). By no means is this the only design. We only claim that this is
feasible and show that it can be used to implement a variety of data-plane
algorithms, which is far beyond a fixed-function switch today. We anticipate
\absmachine machines evolving as data-plane algorithms demand more of
the hardware.

\subsection{Compiling \pktlanguage programs to \absmachine machines}
\label{ss:compiler}
We now consider every target from Table~\ref{tab:templates}, and every
data-plane algorithm from Table~\ref{tab:algos} to determine if the algorithm
can run at line rate on a particular \absmachine machine. We say an algorithm
can run at line rate on a \absmachine machine if every codelet within the
data-plane algorithm can be mapped (\S\ref{ss:code_gen}) to either the stateful
or stateless atom provided by the \absmachine machine. Because stateful atoms
are arranged in a containment hierarchy, we list the \textit{least expressive}
stateful atom/target required for each data-plane algorithm in
Table~\ref{tab:algos}.

We note two lessons for designing programmable
switches from Table~\ref{tab:algos}.
First, atoms supporting stateful operations on a single state
variable are sufficient for several data-plane algorithms. For instance, the
algorithms from Bloom Filter through DNS TTL Change Tracking in
Table~\ref{tab:algos} can be run at line rate using the Nested Ifs atom that
manipulates a single state variable. Second, there are algorithms that need
to update a pair of state variables atomically. One example is CONGA,
whose code we reproduce below:
\begin{verbatim}
  if (p.util < best_path_util[p.src]) {
    best_path_util[p.src] = p.util;
    best_path[p.src] = p.path_id;
  } else if (p.path_id == best_path[p.src]) {
    best_path_util[p.src] = p.util;
  }
\end{verbatim}
Here, \texttt{best\_path} (the path id of the best path for a particular
destination) is updated conditioned on \texttt{best\_path\_util} (the
utilization of the best path to that destination)\footnote{{\tt p.src} is the
  address of the host originating this message, and hence the destination for
the host receiving it and executing CONGA.} and vice versa. These two state
variables cannot be separated into different stages and still guarantee a
packet transaction's semantics. The Pairs atom, where the update to a state
variable is conditioned on a predicate of a pair of state variables, allows us
to run CONGA at line rate.

While the targets in Table~\ref{tab:templates} are sufficient for several
data-plane algorithms, there are algorithms that they can't run at line rate.
An example is CoDel, which cannot be implemented because it requires a square
root operation that isn't provided by any of our targets. One possibility is a
look-up table abstraction that allows us to approximate such
mathematical functions. We leave this exploration to future work.

\textbf{Compilation time:}
Compilation time is dominated by SKETCH's search procedure.  To speed up the
search, we limit SKETCH to search for constants (e.g., for addition) of size up
to 5 bits, given that the constants seen within stateful codelets in our
algorithms are small. Our longest compilation time is 10 seconds when CoDel
doesn't map to a \absmachine machine with the Pairs atom because SKETCH has to
rule out every configuration in its search space.  This time will increase if
we increase the bit width of constants that SKETCH has to search; however,
because the data-plane algorithms themselves are small, we don't expect
compilation times to be a concern.

\subsection{Performance vs. programmability}
\label{ss:perfprog}
While powerful atoms like Pairs can implement more data-plane algorithms, they
have a performance cost.  A more expressive atom incurs longer signal
propagation delays and implies a lower clock frequency or line rate
(the inverse of propagation delay).  To quantify
this intuition, we consider each stateful atom from Table~\ref{tab:templates}
and synthesize a circuit with the lowest possible delay. As we increase the
complexity of the atom, the number of algorithms from Table~\ref{tab:algos}
that it can implement increases (programmability), while at the same time, its
achievable line rate (performance) decreases
(Table~\ref{tab:perfprog}).\footnote{The slightly non-monotonic behavior
between PRAW and IfElseRAW is because the logic synthesis tool is not optimal
and employs many heuristics.} This decrease in line rate can be explained by
looking at the simplified circuit diagrams for the first three atoms
(Table~\ref{tab:circuits}), which show an increase in circuit depth with atom
complexity.

\begin{table}[!t]
  \begin{scriptsize}
  \begin{tabular}{|p{0.08\textwidth}|p{0.12\textwidth}|p{0.09\textwidth}|p{0.09\textwidth}|}
  \hline
  Atom & Min. delay (picoseconds) & Programmability (\# of algorithms implemented by atom) & Performance (Max. line rate in billion pkts/sec) \\
  \hline
  Write & 176 & 1  & 5.68 \\
  \hline
  ReadAddWrite (RAW) & 316 & 2 & 3.16\\
  \hline
  Predicated ReadAddWrite (PRAW) & 393 & 4 & 2.54 \\
  \hline
  IfElse ReadAddWrite (IfElseRAW) & 392 & 5 & 2.55 \\
  \hline
  Subtract (Sub) & 409 & 6 & 2.44 \\
  \hline
  Nested Ifs (Nested) & 580 & 9 & 1.72 \\
  \hline
  Paired updates (Pairs) & 609 & 10 & 1.64 \\
  \hline
  \end{tabular}
\end{scriptsize}
\caption{Programmability increases with more complex atoms,
  but performance decreases.}
\label{tab:perfprog}
\end{table}

\begin{table}[!t]
  \begin{scriptsize}
    \begin{tabular}{|p{0.08\textwidth}|p{0.28\textwidth}|p{0.05\textwidth}|}
  \hline
  Atom & Circuit & Min. delay in picoseconds \\
  \hline
  Write & \includegraphics[width=0.2\textwidth]{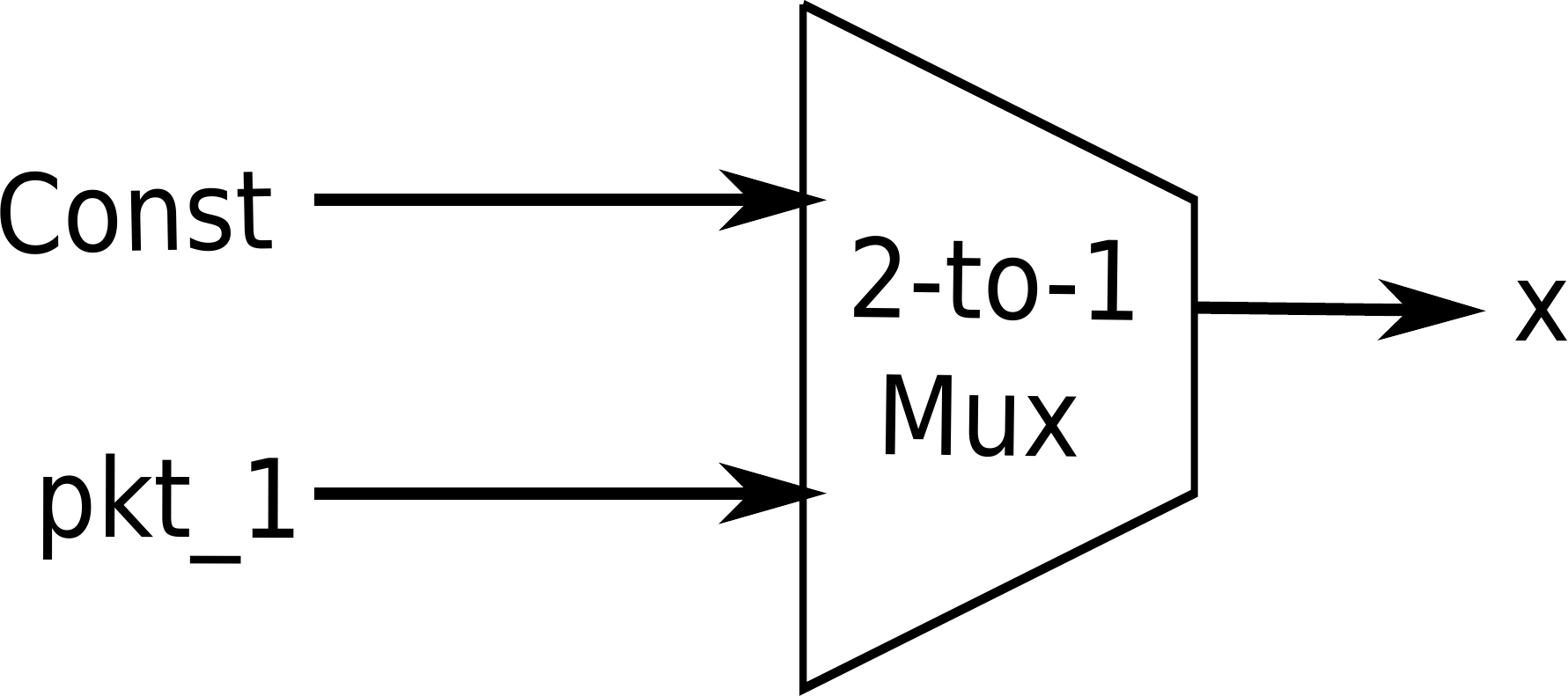} & 176 \\
  \hline
  ReadAddWrite (RAW) & \includegraphics[width=0.2\textwidth]{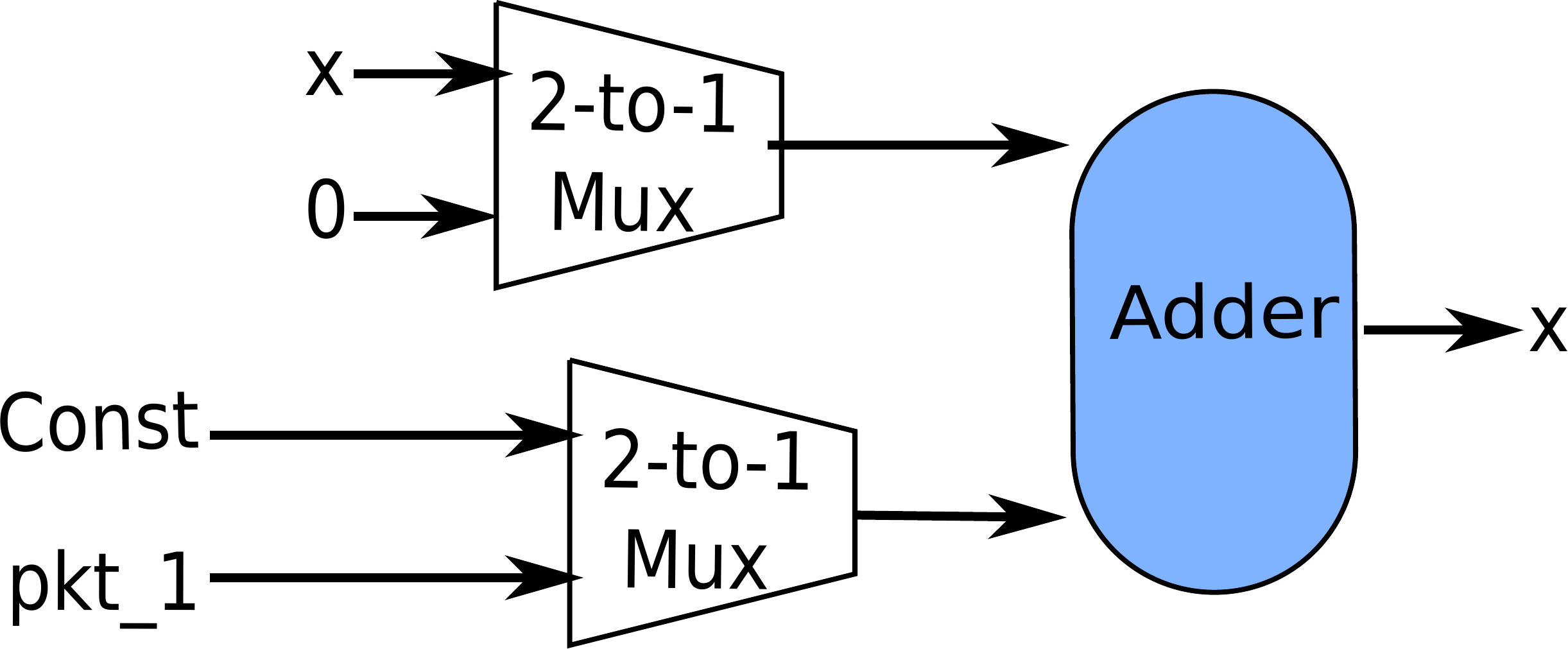} & 316\\
  \hline
  Predicated ReadAddWrite (PRAW) & \includegraphics[width=0.3\textwidth]{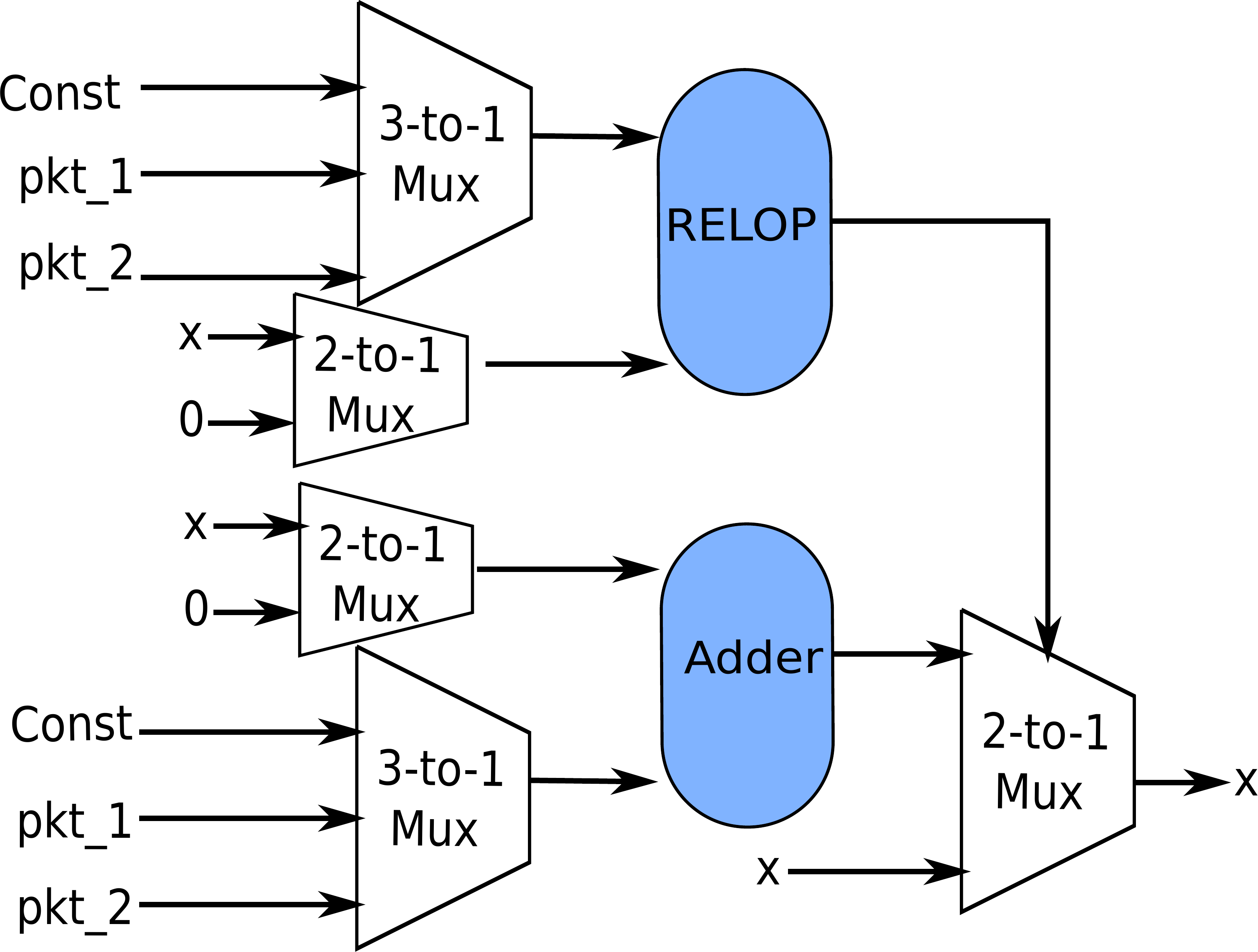}  & 393 \\
  \hline
  \end{tabular}
\end{scriptsize}
\caption{Minimum delay of an atom increases with circuit depth. MUX
stands for a multiplexer, RELOP stands for a relational operation between two
operands.}
\label{tab:circuits}
\end{table}

\section{Related work}
\label{s:related}
\textbf{Abstract machines for line-rate switches:}
NetASM~\cite{netasm} is an abstract machine and intermediate representation
(IR) for programmable data planes that is portable across network
devices---FPGAs, virtual switches, and line-rate switches.  \absmachine is a
machine model for line-rate switches alone, and hence models practical
constraints required for line-rate forwarding that NetASM doesn't. For
instance, \absmachine machines don't permit sharing state between atoms and use
atom templates to limit computations that can happen at line rate. Further,
while NetASM's dataflow framework focuses only on target-independent middle-end
optimizations such as dead-code elimination, the \pktlanguage compiler
implements a compiler back-end for line-rate switches
(\S\ref{ss:code_gen}).

\textbf{Programmable data planes:}
Software data planes such as Click~\cite{click},
RouteBricks~\cite{routebricks}, and Fastpass~\cite{fastpass} are flexible but
lack the performance required for large-scale deployments. Network
Processors~\cite{ixp2800, ixp4xx} (NPUs) were an attempt to bridge the gap.
NPUs are faster than software routers; yet, they remain
\textasciitilde10$\times$ slower than switching chips~\cite{rmt}.

Eden~\cite{eden} provides a programmable data plane using commodity switches by
programming end hosts alone. \pktlanguage targets programmable switches that
increase the scope of programmable data planes relative to an end-host-only
solution. For instance, \pktlanguage permits us to express in-network
congestion control, AQM, and congestion-aware load balancing (CONGA), which are
beyond Eden's capabilities. Tiny Packet Programs (TPP)~\cite{tpp} allow end
hosts to embed small programs in packet headers, which are then executed by the
switch. TPPs are written in a restricted instruction set to facilitate switch
execution; we show that switch instructions must and can be substantially
richer (Table~\ref{tab:templates}) to support stateful data-plane algorithms.

An alternative is to utilize hardware such as 
FPGAs; examples include NetFPGA~\cite{netfpga},
Switchblade~\cite{switchblade}, and Chimpp~\cite{chimpp}.  These designs are
slower than switching ASICs, and are rarely used in production network
equipment. The Arista 7124 FX~\cite{7124fx} is a commercial switch with an
on-board FPGA, but its capacity is limited to 160 Gbits/sec when using the
on-board FPGA---10$\times$ less than the multi-terabit capacities of programmable
switch chips~\cite{xpliant}.

Jose et al.~\cite{lavanya_compiler} focus on compiling P4 programs to
programmable data planes such as the RMT and FlexPipe architectures. Their work
focuses on compiling stateless data-plane tasks such as forwarding and routing,
while \pktlanguage focuses on stateful data-plane algorithms.

\textbf{Packet-processing languages:}
Many programming languages target the network control plane. Examples include
Frenetic~\cite{frenetic}, Pyretic~\cite{pyretic}, and Maple~\cite{maple}.
\pktlanguage focuses on the data plane instead, which requires different
programming constructs and compilation techniques.

Several DSLs target the data-plane. Click~\cite{click} uses C++ for packet
processing on software routers. NOVA~\cite{nova}, packetC~\cite{packetc},
Intel's auto-partitioning C compiler~\cite{intel_uiuc_pldi},
PacLang~\cite{paclang_lang, paclang_partitioner}, and Microengine
C~\cite{microenginec, intel_ixa} target network processors~\cite{ixp2800,
ixp4xx}. \pktlanguage's C-like syntax and sequential semantics are inspired by
these DSLs. However, by targeting line-rate switches, \pktlanguage is more
constrained: e.g., it needs to ensure that the compiled programs can run
at line-rate, hence the language 
forbids loops and includes no synchronization constructs
as there is no shared state in \absmachine machines.

The SNAP system~\cite{snap} programs stateful data-plane algorithms using a
network transaction: an atomic block of code that treats the entire network as
one switch~\cite{onebigswitch} and uses a compiler to translate network
transactions into rules on each switch. SNAP doesn't compile these switch-local
rules into a switch's pipeline. \pktlanguage can be used to compile SNAP's
switch-local rules to an atom pipeline and is an enabler for SNAP and other
network-wide abstractions. FAST~\cite{fast} is another system that provides switch
support and software abstractions for state machines. \absmachine's atoms
support more general stateful processing beyond state machines that enable
a much wide class of data-plane algorithms to be implemented.

\section{Conclusion}
\label{s:conclusion}

This paper presented \pktlanguage, a C-like imperative language that allows
programmers to write packet-processing code using packet transactions, which are
sequential code blocks that are atomic and isolated from other such code
blocks. The \pktlanguage compiler compiles packet transactions to be executed on
\absmachine,
which is a machine model based on programmable line-rate switch
architectures~\cite{flexpipe, xpliant, rmt}. Our results suggest that it is
possible to have both a familiar programming model and line-rate
performance, provided that the algorithm can indeed run at line rate.
Packet-processing languages are still in their infancy; we hope these results
will prompt further work on programming abstractions for packet-processing hardware.

\balance
\bibliographystyle{acm}
\bibliography{paper}

\end{document}